\documentclass[aps,twocolumn,amsmath,showpacs,showkeys,prl]{revtex4-1}

\usepackage{float}
\usepackage{epsfig}
\usepackage{graphicx}
\usepackage{mathptmx}      
\usepackage{latexsym}
\usepackage{natbib}

\begin{document}

\title{Error estimates on Nuclear Binding Energies from Nucleon-Nucleon 
  uncertainties}

\author{R. Navarro P\'erez}\email{rnavarrop@ugr.es}
\affiliation{Departamento de F\'{\i}sica At\'{o}mica, Molecular y
  Nuclear and Instituto Carlos I de F{\'\i}sica Te\'orica y Computacional \\
Universidad de Granada, E-18071 Granada, Spain.}
\author{J.E. Amaro}\email{amaro@ugr.es}
\affiliation{Departamento de F\'{\i}sica At\'{o}mica, Molecular y
  Nuclear and Instituto Carlos I de F{\'\i}sica Te\'orica y Computacional \\
Universidad de Granada, E-18071 Granada, Spain.}
\author{E. Ruiz Arriola}\email{earriola@ugr.es}
\affiliation{Departamento de F\'{\i}sica At\'{o}mica, Molecular y
  Nuclear and Instituto Carlos I de F{\'\i}sica Te\'orica y Computacional \\
Universidad de Granada, E-18071 Granada, Spain.}

\date{\today}

\begin{abstract} 
\rule{0ex}{3ex} Despite great theoretical efforts the NN interaction
can only be determined with a finite precision, implying an error 
upper bound for nuclear masses. We analyze for the first time the problem of
estimating the systematic errors related to the form of the potential
and their impact on nuclear binding. To this end we exploit the concept
of coarse grained interactions to typical nuclear wavelengths. Our
estimate gives an error $\Delta B /A \sim 0.1-0.4 {\rm MeV} $ for the
binding energy per particle and paves the way for {\it ab initio}
calculations tailored to such a precision.
\end{abstract}


\maketitle


Since the early days of Nuclear Physics the NN interaction has played
a major role in the description of the properties of finite nuclei.
While abundant sets of np and pp data have been collected along the
years and accurate theoretical analyses have been carried out since
the
mid-nineties~\cite{Stoks:1993tb,Stoks:1994wp,Wiringa:1994wb,Machleidt:2000ge,Gross:2008ps},
the impact of the NN uncertainties in the Nuclear Many Body Problem
remains an open challenge. This is of utmost importance
as it determines {\it a priori} a lower bound on the inaccuracy of
first principles calculations~\cite{Pieper:2001mp} and might help to
advantageously optimize the computational cost.  While Nuclear binding
energies are experimentally known to high accuracy $\Delta B= 0.01-10
{\rm KeV}$, liquid-drop model inspired mass fit formulae yield a lower
theoretical accuracy $\Delta B= 0.6 {\rm MeV}$ (see
e.g. Refs.~\cite{Toivanen:2008im,1742-6596-267-1-012062} and
references therein). In the present work we face the problem squarely
from the NN side by deducing and propagating two-body systematic
errors to provide a first theoretical {\it a priori} estimate of
binding energy uncertainties.

Error analysis of NN phase-shifts for several partial waves became
first possible when the Nijmegen group~\cite{Stoks:1993tb} carried out
a Partial Wave Analysis (PWA) fitting about 4000 experimental np and
pp data (after rejecting further 1000 of $3\sigma$-mutually
inconsistent data) with $\chi^2/{\rm dof} \sim 1$. The fit fixed the
form of the potential to be an energy dependent square well 
located at a distance of $1.4 {\rm fm}$, a One-Pion-Exchange (OPE) and
Charge-Dependent (CD) contribution starting at $1.4 {\rm fm}$ and a
One-Boson-Exchange (OBE) piece operating below $2-2.5 {\rm
  fm}$. Unfortunately, the required energy dependence becomes messy
for Nuclear Structure calculations.  At present there are a variety of
NN (energy independent) potentials fitting a large body of scattering
data with $\chi^2/{\rm dof} \sim
1$~\cite{Stoks:1993tb,Stoks:1994wp,Wiringa:1994wb,Machleidt:2000ge,Gross:2008ps},
but surprissingly error estimates on potential parameters are not
given. Whereas all these modern potentials share the venerable and {\it
  local} OPE and CD tail and
include electromagnetic effects, the unknown short range components of
these potentials display a variety of forms and shapes, local
potentials~\cite{Stoks:1994wp}, or nonlocal ones implementing angular
momentum dependence~\cite{Wiringa:1994wb}, energy
dependence~\cite{Stoks:1993tb} or linear momentum
dependence~\cite{Stoks:1994wp,Machleidt:2000ge,Gross:2008ps}. While in
principle $p-$, $L-$ and $E-$non-localities are on-shell equivalent
(see e.g. Ref.~\cite{Amghar:1995av} for a proof in a $1/M_N$
expansion) they reflect truely different physical effects and
generally one should consider them as independent quantities; any
specific choice is biased and hence becomes a source of systematic
errors.

We distinguish as usual in error analyses two sources of
uncertainties: statistical errors stemming from the data uncertainties
for a {\it fixed} form of the potential, and systematic errors arising
from the {\it different} most-likely forms of the potentials. Clearly,
the total uncertainty corresponds to adding both in quadrature. In
what follows it is adventageous to take the viewpoint of considering
any of the different potentials as an independent but possibly {\it
  biased} way to measure the scattering amplitudes and/or
phase-shifts.  Because the biases introduced in all single potential
are independent on each other, a randomization of systematic errors
makes sense. Thus, the overall spread between the various
phenomenological models with $\chi^2 / {\rm dof} \sim 1$ provides the
scale of the uncertainty.

\begin{figure}[ht]
\begin{center}
\epsfig{figure=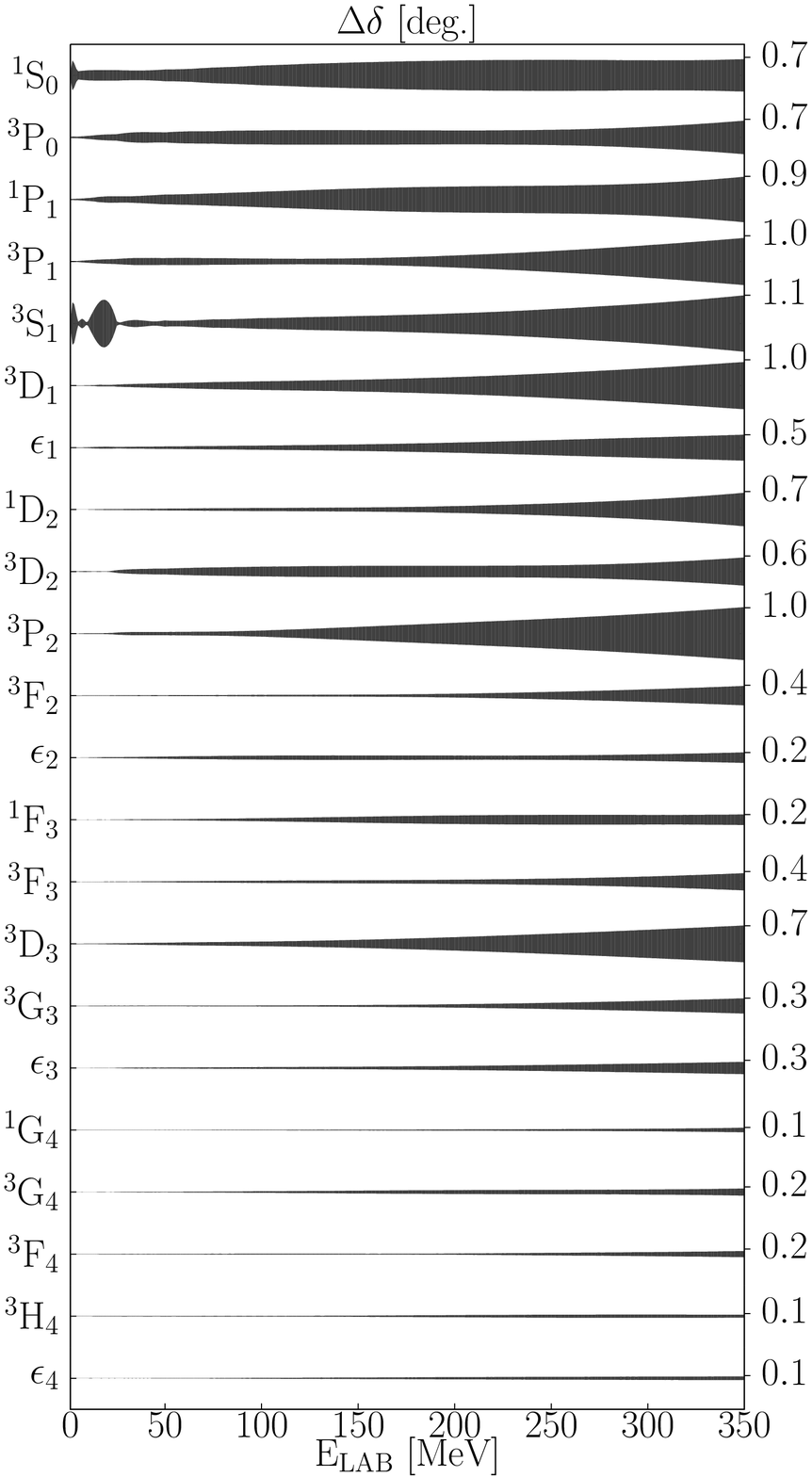,height=12cm,width=8cm}
\end{center}
\caption{Absolute errors (in degrees, right axis) for partial wave phase shifts
  with $J \le 4$ (left axis) due to 7 different potentials fitting scattering data
  with $\chi^2/{\rm dof} \sim
  1$~\cite{Stoks:1993tb,Stoks:1994wp,Wiringa:1994wb,Machleidt:2000ge,Gross:2008ps}
  as a function of the LAB energy (in MeV).}
\label{fig:errors-ps}
\end{figure}

In Fig.~\ref{fig:errors-ps} we show the absolute (mean-square) errors
for np partial wave phase shifts due to the different potentials
fitting scattering data with $\chi^2/{\rm dof} \sim
1$~\cite{Stoks:1993tb,Stoks:1994wp,Wiringa:1994wb,Machleidt:2000ge,Gross:2008ps}
as a function of the LAB energy. As one naturally expects the
uncertainties grow with energy and decrease with the relative angular
momentum which semiclassically corresponds to probing an impact
parameter $ b \sim (L+1/2)/ p $, with $p= \sqrt{M_N E_{\rm LAB} /2}$
the CM momentum, making peripheral waves to be mostly determined from
OPE. These analyses stop at the pion production threshold so that one
probes distances till $b_{\rm min} \sim 1/\Lambda = 0.5 {\rm fm}$ with
$ \Lambda=\sqrt{m_\pi M_N}$. Generally, the PWA statistical
errors~\cite{Stoks:1993tb} turn out to be {\it smaller} than the
systematic bands displayed in Fig.~\ref{fig:errors-ps}.  This
counter-intuitive result relies not only on the specific forms of
potentials which treat the mid-- and short-range behaviour of the
interaction differently but also on the fact that the fits are mainly
done to scattering amplitudes rather than to the phase-shifts
themselves. Our purpose is to quantify the impact of uncertainties in
Fig.~\ref{fig:errors-ps} on Nuclear Binding energies.

The most direct way of analyzing binding energy uncertainties from
randomized systematic errors would be to undertake large scale {\it ab
  initio} calculations using the different forms of the set of $N$
two-body potentials, say $V_2^{(i)}$ with $i=1, \dots N$, yielding
$B^{(i)}(A)$ whence a mean $\bar B(A)$ and a standard deviation
$\Delta B(A)$ can be constructed.  For instance, the triton binding
energy obtained by Faddeev calculations is $ 8.00, 7.62, 7.63, 7.62,
7.72 $ MeV for the CD Bonn~\cite{Machleidt:1995km}, Nijm-II, Reid93,
Nijm-I and AV18~\cite{Friar:1993kk} respectively. More recently, the
covariant spectator model has produced the closest binding energy
$8.50$ MeV to experiment precisely when the NN $\chi^2/{\rm dof}$ is
smallest. This yields in all $B_3 = 7.85(34) {\rm MeV}$ (exp. $B_3 =
8.4820(1) {\rm MeV}$) i.e. $\Delta B_3/3 = 0.11 {\rm MeV}$.  Of course,
in doing so even for the triton or the $\alpha$-particle there is
typically a flagrant need for three-body interactions which account
for the missing $1 {\rm MeV}$ and $4 {\rm MeV}$ to the binding energy
respectively. On the other hand, the definition of the three- and
higher-body interaction depends on the two body potential, so any
uncertainty in the two-body interaction will {\it carry over} to the
three-body interaction. Thus, even if we fix it say in the $A=3$
system, there will always be a residual uncertainty in the $A+1=4$
calculation.  Thus, estimating the two-body uncertainty provides a
lower bound on the total uncertainty if the correlations between the
two- and three-body forces are ignored. The argument generalizes trivially to
any $A-$body interactions and $A+1-$nuclei.

Unfortunately, the procedure outlined above of using different
potentials stops beyond the $A=4$ nucleus, due to computational and
theoretical difficulties related to the form of the potential.  From
an {\it ab initio} viewpoint, only Monte Carlo calculations may go up
to $A=10$ when potentials are fixed to be $r-$dependent with a
nonlocality in terms of the relative angular momentum
operator~\cite{Lagaris:1981mm,Wiringa:1984tg}. For that good reason
the Argonne potential saga has been constructed sticking to this
representation and culminating in the AV18
potential~\cite{Wiringa:1994wb}, an updated version of AV14 containing
charge-independence-breaking (CIB) terms and a complete
electromagnetic interaction and fitted directly to 4301 $pp$ and $np$
data from the 1993 Nijmegen partial-wave analysis~\cite{Stoks:1993tb}
with the requested $\chi^2 / {\rm dof} \sim 1$. The AV18 and AV18+UIX
have become standard Hamiltonians for {\it ab initio} calculations of
light nuclei~\cite{Pieper:2001mp} and dense
matter~\cite{Akmal:1997ft}.  Note that even if statistical errors on
potential parameters would have been estimated in Ref.~\cite{Wiringa:1994wb},
the question on the systematic errors remains. We motivate
below an approximate method to address these issues.

One of the outstanding features of the AV18-potential is the presence
of the short distance core in the central part, $V_C(r)$, for
distances below $a_{\rm core}=0.5 {\rm fm}$, which demands sizeable
and fine-tuned short distance
correlations~\cite{Pieper:2001mp}. However, for a closed-shell nucleus
one has schematically,
\begin{eqnarray}
\langle V_2 \rangle_A = \frac{A(A-1)}{2} \int d^3 r \, P_2 (r) \, V_C (r) \, , 
\end{eqnarray}  
where $P_2 (r)$ is the probability of finding two particles at a
distance $r$ that turns out to be fairly independent of the particle
number. In particular, for $r \lesssim a_{\rm core} $ one has $V_C (r)
\gg B/A $ and one is left with a two-body problem with $A-2$
spectators; in the classically forbidden region an exponential
suppression, $ P_2 (r) \sim \exp( -2 \int_r^{a_{\rm core}} dr \sqrt{ 2
  \mu V_C (r)}) $ is expected semiclassically. Thus the contribution
from the core is small, precisely in the region where the NN force is
not well determined from the PWA probing $r \ge b_{\rm min}$. For our error estimate we propose to
side-step this core complication by introducing a coarse grained
potential where the cancellation of the product $P_2(r) V_C(r)$ comes
from a vanishing potential below $a_{\rm core}$.

The previous argument was suggested long ago by Afnan and
Tang~\cite{Afnan:1968zj} who realized that for $A=3,4$ systems the
relevant NN-scattering energies do not probe the core expliticly.
Soft core potentials, fitted to NN low partial waves up to $E_{\rm LAB}=100$MeV,  
provided reasonable binding energies. Furthermore the hard core can be
made into a soft core by introducing (linear-momentum) nonlocalities
in terms of the kinetic energy operator by a unitary phase-preserving
transformation~\cite{Neff:2002nu}. Actually, this is the physics
behind the so-called $V_{\rm low k}$ potentials based in the
definition of an effective truncated Hilbert space below a given
cut-off $\Lambda \sim \sqrt{M_N
  m_\pi}$~\cite{Bogner:2003wn,Bogner:2009bt}.  Recently, we have shown
how a similar idea can be implemented in coordinate space using a
coarse grained potential~\cite{Perez:2011fm,Perez:2012qf}, i.e. an
average potential over a given wavelength resolution $\Delta r \sim
b_{\rm min}$ ; that means specifying the potential information in a
finite number of points.  The form of the potential is not important
but calculations become simple by taking delta-shells in the
region below $3 {\rm fm}$. For the partial wave $^{2S+1}(l,l')_J$
the potential reads 
\begin{eqnarray}
V^{JS}_{l,l'}(r) = \frac{1}{2\mu}\sum_{n=1}^N  (\lambda_n)^{JS}_{l,l'} 
\delta(r-r_n) \, ,  \qquad r \le r_c \, , 
\label{eq:ds-pot}
\end{eqnarray}
with $\mu$ the reduced pn-mass and $r_c= 3 {\rm fm}$.  In
practice  $N \le 5 $ for any given partial
wave. For $r > 3 {\rm fm}$ we use the customary CD OPE+electromagnetic
interactions. The main novelty is a determination of the, so far,
{\it missing errors} in the potential parameters (in this case
$(\lambda_n)^{JS}_{l,l'} $) from the uncertainties depicted in
Fig.~\ref{fig:errors-ps} and corresponding to all $\chi^2 / {\rm dof}
\sim 1 $
fits~\cite{Stoks:1993tb,Stoks:1994wp,Wiringa:1994wb,Machleidt:2000ge,Gross:2008ps}. For
instance, we found~\cite{Perez:2012qf} by using the $^3S_1$, $^3D_1$
and $E_1$ uncertainties of Fig.~\ref{fig:errors-ps} that for the
deuteron $\Delta B_{^2{\rm H}} /2 = 0.1 {\rm MeV}$ compared to the experimental 
 $\Delta B_{^2{\rm H}} /2 = 0.005 {\rm KeV}$. 

In our previous calculation~\cite{Perez:2011fm}, we showed how our
approach is competitive not only as a way of determining the phase
shifts but also with more sophisticated approaches to Nuclear
Structure~\cite{Neff:2002nu}. This was checked with oscillator wave
functions in the case of $^4 {\rm He}$, $^{16}{\rm O}$ and $^{40}{\rm
  Ca}$ which reproduces experiment at the $20-30\%$-level provided the phase-shifts are fitted to about $100
{\rm MeV}$~\cite{Perez:2011fm}. This is a
tolerable accuracy as we just intend to make a first estimate on
the systematic uncertainties and then compute the change in the
binding energy from the simple formulas,
\begin{eqnarray}
\Delta B_{^3{\rm H}} &=&\langle \Delta V_2 \rangle_{^3{\rm H}} = 3 \langle 1s | \frac12 \left( \Delta V_{^1S_0} + \Delta V_{^3S_1} \right) | 1s \rangle  \, , 
\label{eq:db_3}\\ 
\Delta B_{^4{\rm He}} &=&\langle \Delta V_2 \rangle_{^4{\rm He}} = 6 \langle 1s | \frac12 \left( \Delta V_{^1S_0} + \Delta V_{^3S_1} \right) | 1s \rangle \, , 
\label{eq:db_4} 
\end{eqnarray}
where $ |1s \rangle $ is the Harmonic oscillator relative wave
function with the corresponding $b-$ oscillator parameter reproducing
the physical charge radius. The numbers in front are Talmi-Moshinsky
coefficients and correspond in this particular case to the number of
pairs interacting through a relative s-wave. Errors are computed by
adding individual contributions $ (\Delta \lambda_n)^{JS}_{l,l'} $
from Eq.~(\ref{eq:ds-pot}) in quadrature. By propagating the PWA errors in
Eq.~(\ref{eq:db_3}) we find $\Delta B_3/3 = 0.07-0.085 {\rm MeV}$
depending on the fitting cut-off LAB energy, 100-350 MeV respectively,
in good agreement with the Faddeev estimates given above. For the
$\alpha-$particle Eq.~(\ref{eq:db_4}) yields $\Delta B_4/4 =
0.10-0.13 {\rm MeV}$.  Along the lines of Ref.~\cite{Perez:2011fm} we
also find $\Delta B_{^{16}{\rm O}}/16 = 0.26 {\rm MeV}$ and $\Delta
B_{^{40}{\rm Ca}}/40 = 0.32 {\rm MeV}$.

\begin{table*}[htb]
\begin{tabular}{|c|c|c|c|c|c|c|c|c|c|c|}
\hline 
Parameter  & $\epsilon_V $  &  $\epsilon_V $ (AV18)& $^2{\rm H}$  & $^2{\rm H}$ (AV18)&  $^3{\rm H}$ (100) & $^3{\rm H}$ (350) & $^3{\rm H}$ (AV18)& $^4{\rm He}$ (100) & $^4{\rm He}$ (350)
& $^4{\rm He}$ (AV18)  \\
\hline 
$m_N$          & -0.0445  &-0.0444  &-0.3997   & -0.3960  &-1.0030   &-1.1606    & -0.9797    &-1.6820   &-1.9973   &  -2.1945   \\
$m_N +\delta_N $     & -0.1602  &-0.1612   &-0.7986  & -0.7928  &-2.7817   & -3.8063   &-2.6560     &-5.2395  & -7.2887  & -6.5230 \\
$\delta_\Delta $    &  0.1350  & 0.1359  & 0.4641  & 0.4609   &  2.0690  &  3.0778  &1.9504    &  4.1380  & 6.1555    & 5.0380   \\
$m_\pi $ (OPE)   &  -0.0051  &-0.0051   &0.0289   & 0.0623  & 0.0140 & -0.1676     &  0.0730  & 0.0280   &-0.1461  & 0.1045    \\
$m_\pi $ (+TPE-s) &   0.0706  & 0.0705   &0.2665  &  0.2981 & 1.1205  & 1.1770    &  1.1055   &2.2409     &  2.3540 & 2.8105   \\
$m_\pi $ (+TPE-L) &   --        & --  &  0.2683 & 0.2999  & 1.1229  &  1.1812   & 1.1025   &2.2459   &2.3623  & 2.7830    \\
$m_V $    & -0.4757   &-0.5079   &-1.8692   &-1.8571   & -7.9517  & -10.3458  & -7.6743   & -15.9034   &-20.6916   &-19.9870    \\
\hline
\end{tabular}
\caption{\label{tab:change} Changes of ground state energies
  of lightest nuclei $A=2,3,4$ in MeV for the coarse-grained
  potentials with oscillator shell model  when 
  $^1S_0$ and $^3S_1$ phase shifts are fitted up to LAB
  energies $E_{\rm LAB}= 100 {\rm MeV}$ and $E_{\rm LAB}= 350 {\rm
    MeV}$ used in this work compared to the {\it ab initio} Monte
  Carlo  calculations~\cite{Flambaum:2007mj} (labeled as AV18 here and using their notation) . The changes correspond to the
  variation of the total energy when the mass parameter is either
  increased or decreased by $1\%$. For instance the label $m_N$
  corresponds to  $ B( 1.01 m_N )- B(0.99 m_N)$.}
\end{table*}

A simple estimate on the impact of errors due to the two
body interaction uncertainty can be done using  Skyrme effective interactions 
(for a
review see \cite{Bender:2003jk}) 
\begin{eqnarray}
\frac{\Delta B}{A} =   
\frac{3}{8 A} \Delta t_0 \, \int d^3 x \, \rho (x)^2 \,   , 
\end{eqnarray}
where we get $t_0 = (\pi/\mu) \sum_n r_n^2 (\lambda_{^1S_0,n } +
\lambda_{^3S_1,n }) $. Using the two-body interaction of
Eq.~(\ref{eq:ds-pot}) and propagating errors we get $t_0 = -0.92(1)
{\rm GeV} {\rm fm}^3$. As a check on the size of $t_0$ we note that
from a fit to the equation of state used by the Trento
group~\cite{Gandolfi:2009nq} at low densities we obtain a value $t_0
\sim - 0.9(1) {\rm GeV} {\rm fm}^3 $, whereas a coarse graining of NN
interactions in CM momentum space down to $\Lambda\sim 0.3 {\rm GeV}$
gives a compatible value, $t_0 \sim -4
\pi^2/(M_N\Lambda)$~\cite{Arriola:2010hj}.  For nuclear matter at
saturation, $\rho_0= 0.17 {\rm fm}^{-3}$, our $\Delta t_0= 10 
{\rm MeV} {\rm fm}^3$ implies
\begin{eqnarray}
\frac{\Delta B}{A} = \frac{3}{8} \Delta t_0 \rho_0  = 0.6 {\rm MeV}  \, . 
\end{eqnarray}
We may implement finite size effects  by using 
a Fermi-type  shape for the matter density 
$
\rho(r) =   C/(1+e^{(r-R)/a})
$
with  $R=r_0 A^{\frac13}$ and $r_0=1.1 {\rm fm}$ and $a=0.7 {\rm fm}$ and normalized to the total number of particles $A = \int d^3 x \rho(x)$ we get 
\begin{eqnarray}
\Delta B_A/A =  0.1-0.4 {\rm MeV} \, , 
\end{eqnarray}
which depends on the value of $A$ for $4 \le A \le 208$.  

\begin{figure}[ht]
\begin{center}
\epsfig{figure=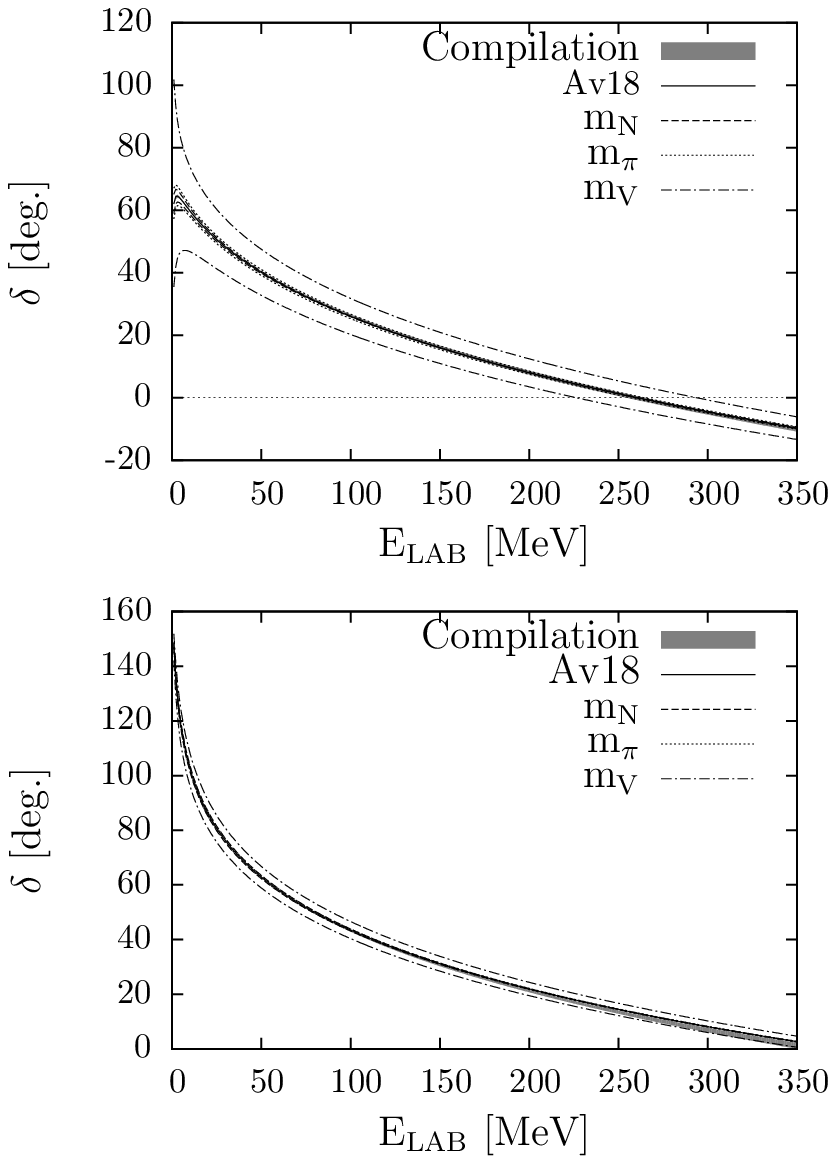,height=10cm,width=6cm}
\end{center}
\caption{AV18 phase shifts in the $^1S_0$ (upper panel) and $^3S_1$
  (lower panel) partial waves as a function of the CM momentum when
  the parameters are varied as specified in
  Ref.~\cite{Flambaum:2007mj}. We also depict the band corresponding
  to the spread of values obtained with 7 high quality
  potentials~\cite{Stoks:1993tb,Stoks:1994wp,Wiringa:1994wb,Machleidt:2000ge,Gross:2008ps}
  containing One-Pion-Exchange (OPE) and Charge Dependence (CD) tail
  and fit scattering data with $\chi^2/{\rm dof} \sim 1$.}
\label{fig:varied-ps}
\end{figure}

One may reasonably doubt that the core effects and the corresponding
short distance correlations can be reliably monitored by a simple
shell model calculation as given by Eq.~(\ref{eq:db_3}) and
Eq.~(\ref{eq:db_4}) for $^3 {\rm H}$ and $^4 {\rm He}$ respectively.
We show now that for the purpose of error estimate this is however not
so. In order to check this we take advantage of a recent analysis
using the AV18-potential~\cite{Flambaum:2007mj} where the dependence
of nuclear binding on the potential parameters is analyzed in much
detail for mass numbers $A=2, \dots 8$. Relative $1\%$ variations of
two-body potential parameters are considered yielding changes in
binding energies in the range of $0.1-20 {\rm MeV}$ for $^4{\rm
  He}$. This is a much larger range than our estimated errors.  We
note here that the largest sensitivity is on the short range
potential, whereas the pion mass variation in the OPE piece yields a
tiny effect. We may test our strategy by proceeding as follows.  For
any variation of the AV18-potential parameters there is a
corresponding change in the phase shifts. In Fig.~\ref{fig:varied-ps}
we show as an illustration the changes in the most important $^1S_0$
and $^3S_1$ waves due to several changes in parameters as explained in
Ref.~\cite{Flambaum:2007mj}. As we see these changes are indeed larger
than the systematic errors depicted in Fig.~\ref{fig:errors-ps}, that
is enough for our purposes.  Given this variation we then readjust our
coarse grained potential, Eq.~(\ref{eq:ds-pot}), as to reproduce such
a change and then use Eqs.~(\ref{eq:db_3},\ref{eq:db_4}). We show our
results in table~\ref{tab:change} depending on the fitted maximal
$E_{\rm LAB}$.  The disagreement of our results with those of AV18 in
the case of changing the pion mass in the OPE potential
is not very important since the net result is rather small anyhow and
the pion mass is well known. We note, however, a larger sensitivity of
the AV18 potential with respect to the short distance variations which
is the relevant aspect for our error analysis.  In other words, our coarse
grained calculation reproduces {\it ab initio} $\Delta B$'s at the $20
\%$ accuracy.

We summarize our points. Nuclear Binding energies are a crucial test
for the Nuclear Many Body Problem. While first principles calculations
are hampered by computational difficulties, we note that nuclear force
uncertainties may have a useful impact on these calculations. The
present theoretical estimates are in the range $\Delta B/A \sim
0.1-0.4 {\rm MeV}$ exceeding two or three orders of magnitude the
avalaible precision of {\it ab initio} Monte Carlo calculations
achieved up to date for $A \le 10$.  The prospective of increasing the
particle number keeping the computational cost provides strong reasons
to go beyond NN uncertainties as done here and to consider also errors
in 3N and 4N forces. Finally, one should keep in mind that agreement
between theory and experiment could also be declared even when the
theory is less precise than the experiment, as it so frequently
happens in Nuclear Physics, provided of course both error bands
overlapp.


{\it 
This work is supported by Spanish DGI
  (grant FIS2011-24149) and Junta de Andaluc{\'{\i}a} (grant FQM225).
  R.N.P. is supported by a Mexican CONACYT grant.}


%

\end{document}